%% file: acl_latex.tex
\pdfoutput=1

\documentclass[11pt]{article}

\usepackage[preprint]{acl}

\usepackage{times}
\usepackage{latexsym}

\usepackage[T1]{fontenc}

\usepackage[utf8]{inputenc}

\usepackage{microtype}

\usepackage{inconsolata}

\usepackage{graphicx}

\usepackage{multirow}  
\usepackage{amssymb}
\usepackage{enumitem}
\usepackage{amsmath}
\usepackage{booktabs}  
\usepackage[noabbrev,capitalize]{cleveref}

\usepackage{colortbl}
\newcommand{\headercolor}{\rowcolor{gray!15}}

\input{math_commands.tex}

\usepackage{tcolorbox}
\usepackage{xspace}

\newtcolorbox[list inside=prompt,auto counter,number within=section]{prompt}[1][]{
    colbacktitle=black!60,
    coltitle=white,
    fontupper=\footnotesize,
    boxsep=5pt,
    left=0pt,
    right=0pt,
    top=0pt,
    bottom=0pt,
    boxrule=1pt,
    #1,
}

\newcommand{\Ours}{DecoupleSearch\xspace}
\usepackage{multicol}
\newcommand{\symboletongyi}{\raisebox{0pt}{~\includegraphics[scale=0.012]{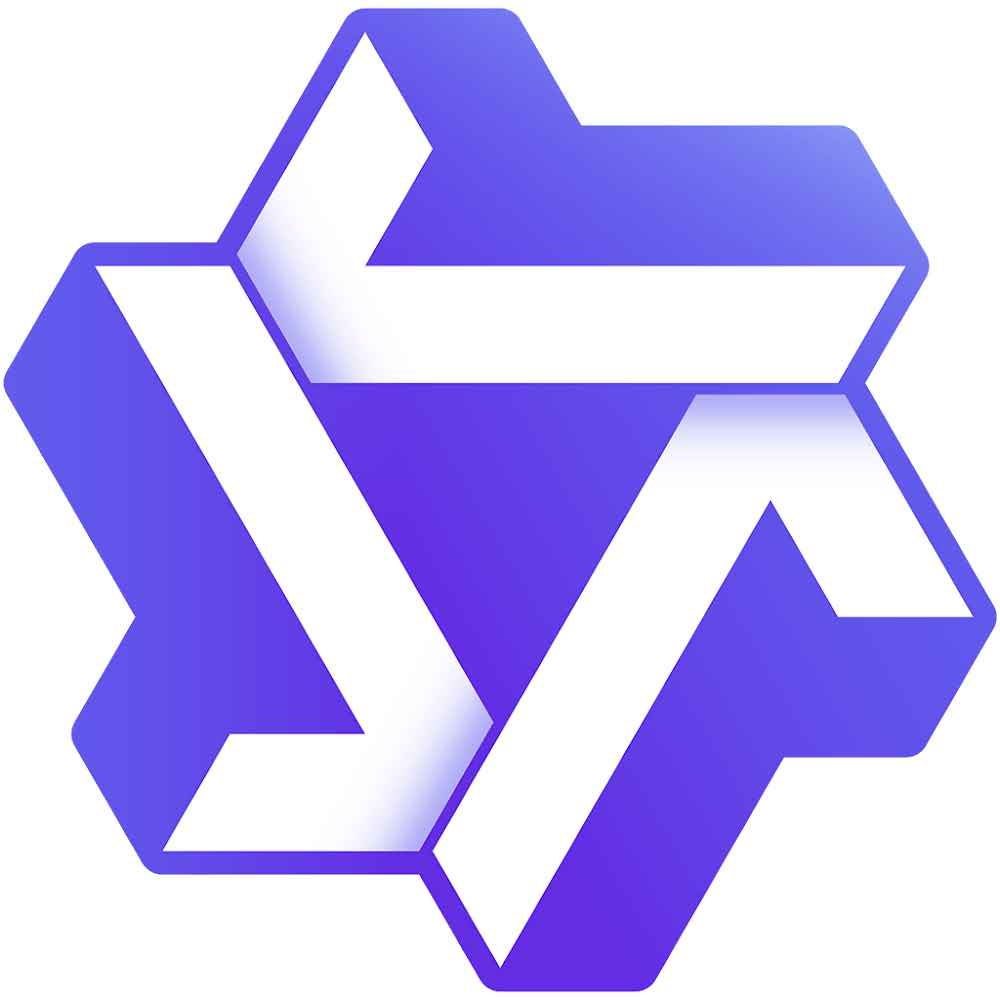}}~}
\newcommand{\huggingface}{\raisebox{-1.5pt}{\includegraphics[height=1.05em]{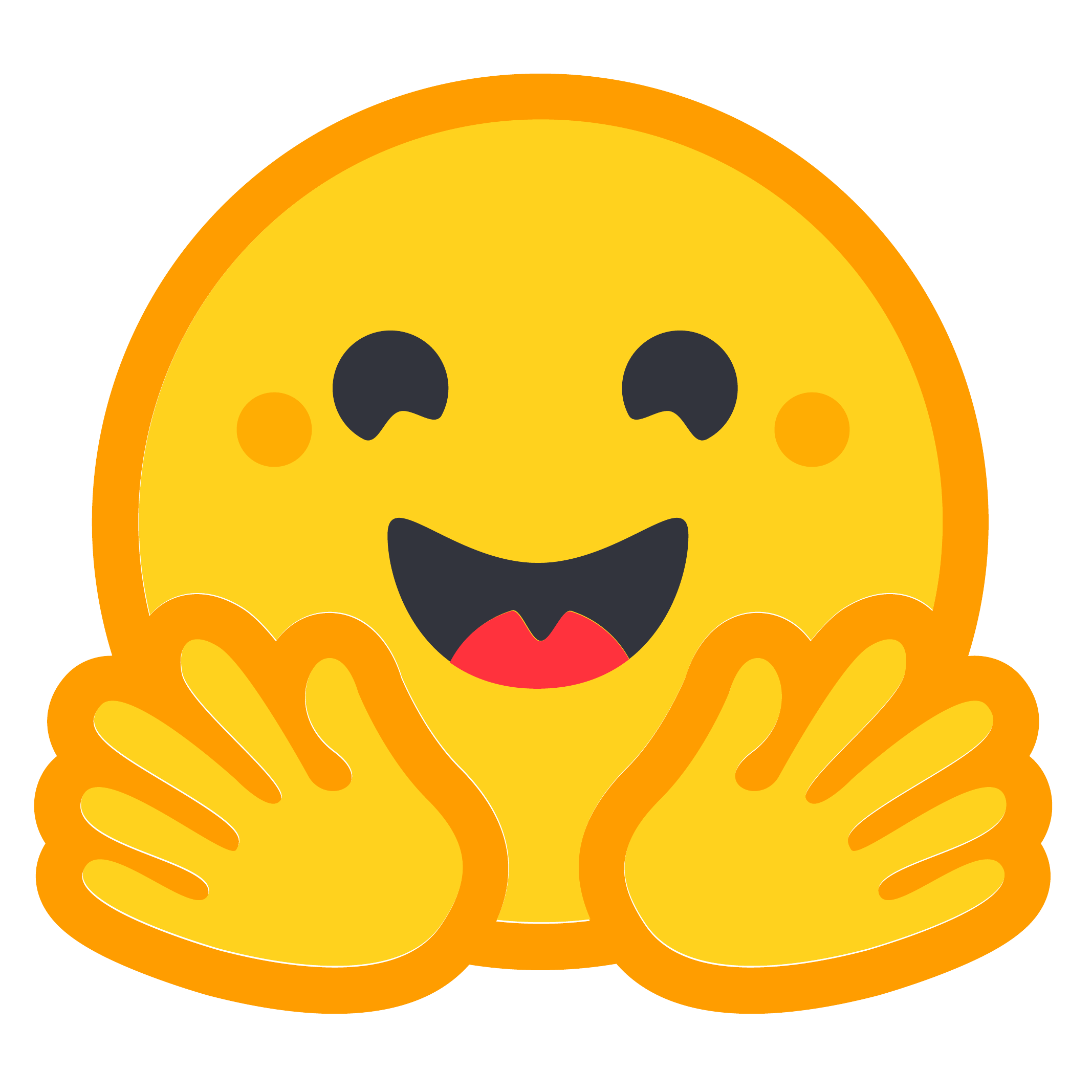}}\xspace}
\newcommand{\hfdataset}{\raisebox{-1.5pt}{\includegraphics[height=1.05em]{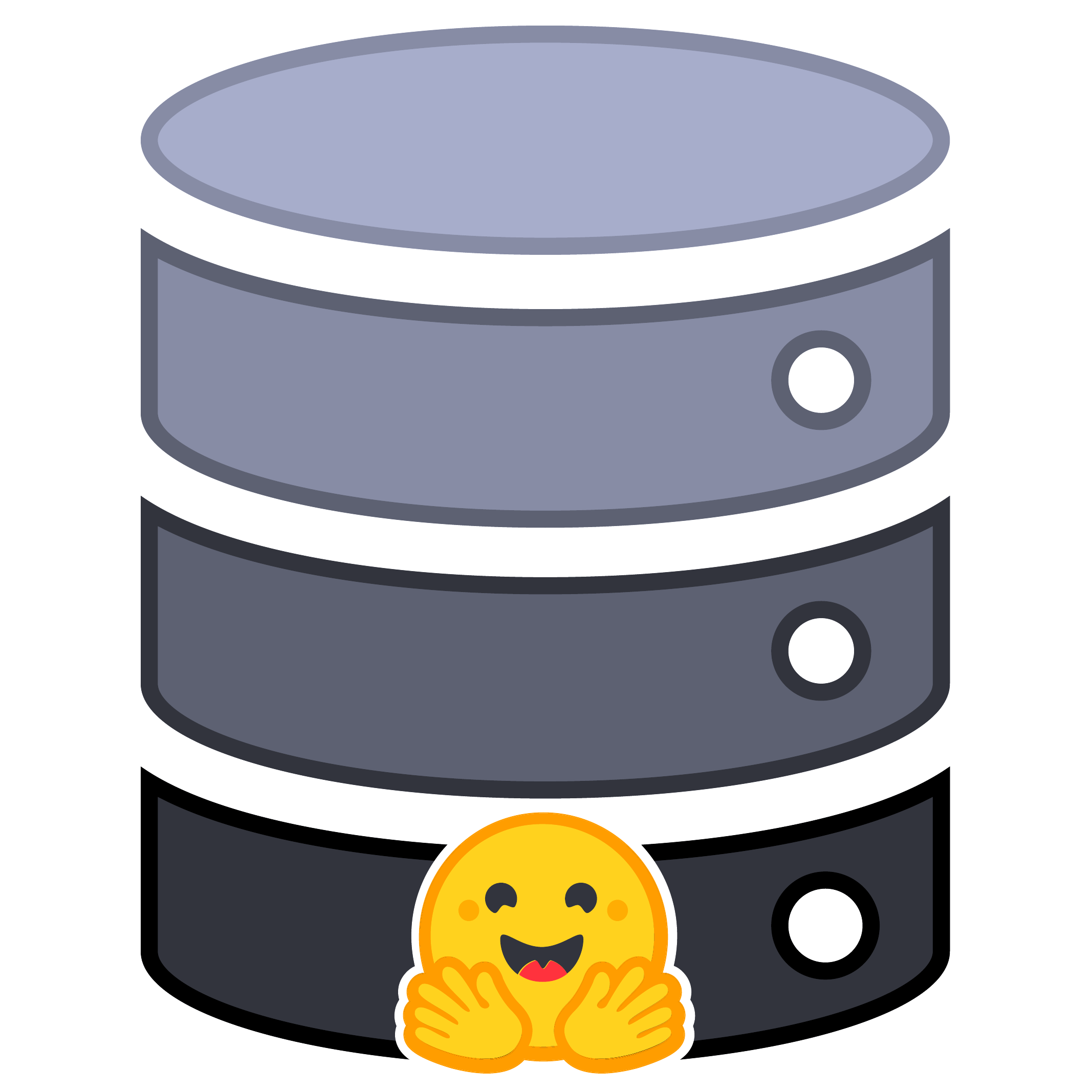}}\xspace}
\newcommand{\github}{\raisebox{-1.5pt}{\includegraphics[height=1.05em]{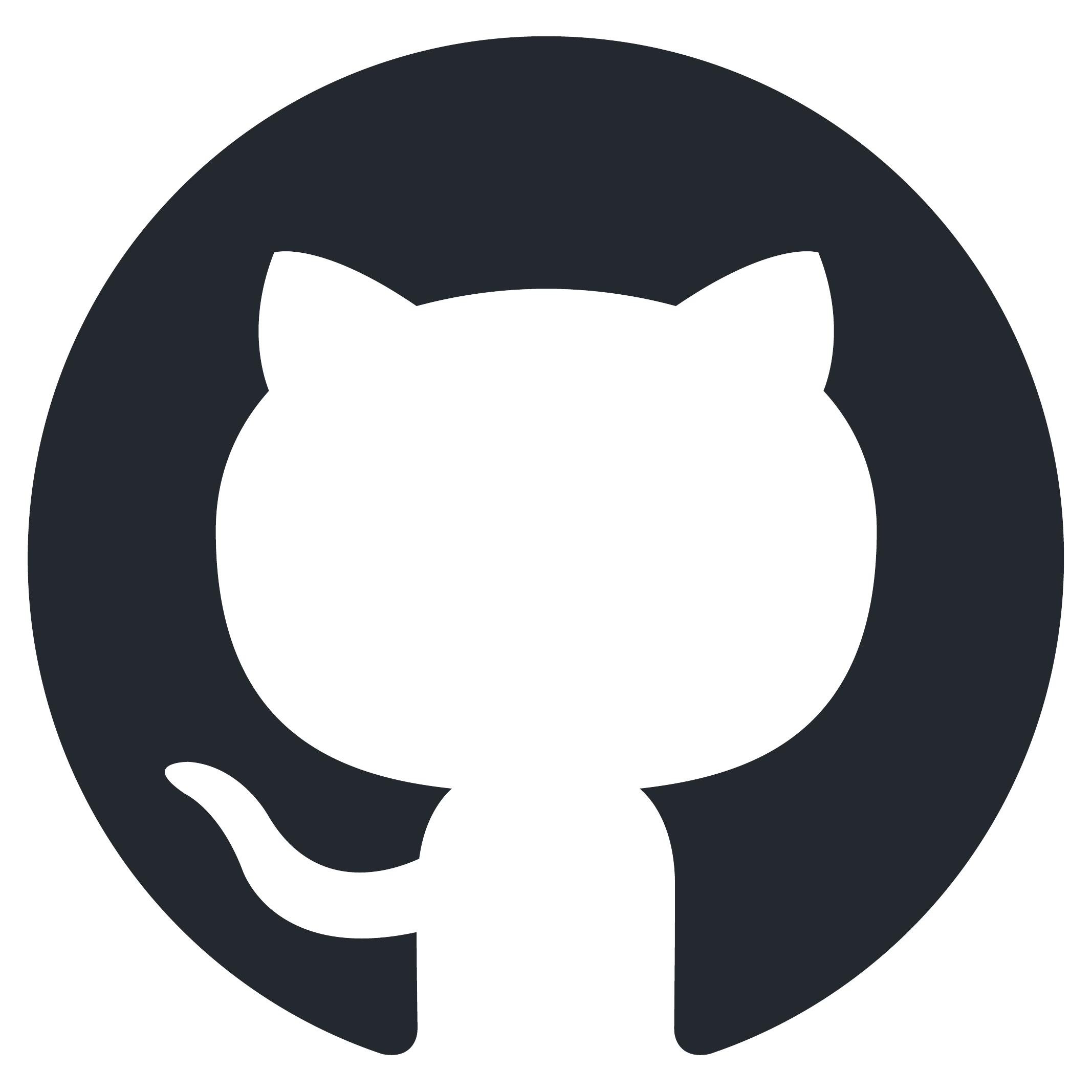}}\xspace}
\newcommand{\project}{\raisebox{0pt}{\includegraphics[height=1.0em]{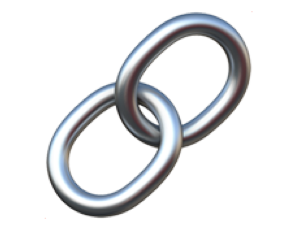}}\xspace}

\title{\Ours: Decouple Planning and Search via\\ Hierarchical Reward Modeling}

\author{%
Hao Sun, Zile Qiao\thanks{Corresponding Author.}, Bo Wang, Guoxin Chen, Yingyan Hou,\\
\textbf{Yong Jiang, Pengjun Xie, Fei Huang, Yan Zhang\textsuperscript{$*$}}\\
Tongyi Lab\symboletongyi, Alibaba Group \\
\vspace{.5em}\project\href{https://sunhaonlp.github.io/DecoupleSearch/}{Homepage} 
\vspace{.5em}\huggingface \href{https://huggingface.co/collections/sunhaonlp/decouplesearch-6860d78abbe29b43f6c347d4}{Model}
\vspace{.5em}\hfdataset \href{https://huggingface.co/datasets/sunhaonlp/DecoupleSearch_dataset}{Datasets} 
\vspace{.5em}\github \href{https://github.com/sunhaonlp/DecoupleSearch}{Code}
}

\begin{document}
\maketitle
\begin{abstract}
Retrieval-Augmented Generation (RAG) systems have emerged as a pivotal methodology for enhancing Large Language Models (LLMs) through the dynamic integration of external knowledge. To further improve RAG's flexibility, Agentic RAG introduces autonomous agents into the workflow. However, Agentic RAG faces several challenges:
(1) the success of each step depends on both high-quality planning and accurate search,
(2) the lack of supervision for intermediate reasoning steps, and
(3) the exponentially large candidate space for planning and searching.
To address these challenges, we propose \textbf{\Ours}, a novel framework that decouples planning and search processes using dual value models, enabling independent optimization of plan reasoning and search grounding. Our approach constructs a reasoning tree, where each node represents planning and search steps. We leverage Monte Carlo Tree Search to assess the quality of each step. During inference, Hierarchical Beam Search iteratively refines planning and search candidates with dual value models. Extensive experiments across policy models of varying parameter sizes demonstrate the effectiveness of our method.
\end{abstract}

\input{Section/Introduction}
\input{Section/Background}
\input{Section/Methodology}
\input{Section/Experiment}
\input{Section/Related_work}
\input{Section/Conclusion}
\input{Section/Limitations}
\input{Section/Ethics}


\bibliography{anthology,custom}

\clearpage

\appendix
\onecolumn
\input{Section/Appendix}

\end{document}

%% file: math_commands.tex
\usepackage{amsmath,amsfonts,bm}

\def\eqref#1{equation~\ref{#1}}

\def\1{\bm{1}}

\DeclareMathAlphabet{\mathsfit}{\encodingdefault}{\sfdefault}{m}{sl}
\SetMathAlphabet{\mathsfit}{bold}{\encodingdefault}{\sfdefault}{bx}{n}



%% file: Section/Introduction.tex
\section{Introduction}
Large Language Models (LLMs) \cite{Taylor-arxiv-2022-Galactica, chowdhery2022palm, zhao2023survey} have demonstrated remarkable performance across a wide range of downstream tasks \cite{xia2024evaluating, yamauchi2023lpml, imani2023mathprompter, lewkowycz2022solving}. Despite these advancements, LLMs remain susceptible to generating responses that include hallucinated facts \cite{ji2023survey, shuster2021retrieval, zhang2023sac3}, undermining their reliability. To address this challenge, Retrieval-Augmented Generation (RAG) has been proposed, integrating external knowledge to enhance the generation process \cite{ram2023context, shi2023replug, rashkin2021measuring, gao2022rarr, bohnet2022attributed, menick2022teaching}.

While RAG systems have led to significant improvements, they still face important limitations. These systems rely on static workflows and struggle to effectively handle multi-step reasoning or complex tasks. A promising solution to these limitations is Agentic Retrieval-Augmented Generation (Agentic RAG), which introduces autonomous AI agents into the RAG pipeline \cite{asai2023self, yu2024auto, chen2024mindsearch, li2025search}. In this framework, the reasoning process typically involves two phases: planning and search. During the planning phase, the agent analyzes the current reasoning process and determines which information is still required. In the search phase, the agent generates search queries to retrieve relevant external documents. These phases alternate iteratively until a final answer is produced.

\begin{figure}[t]
    \centering
    \includegraphics[width=.99\linewidth]{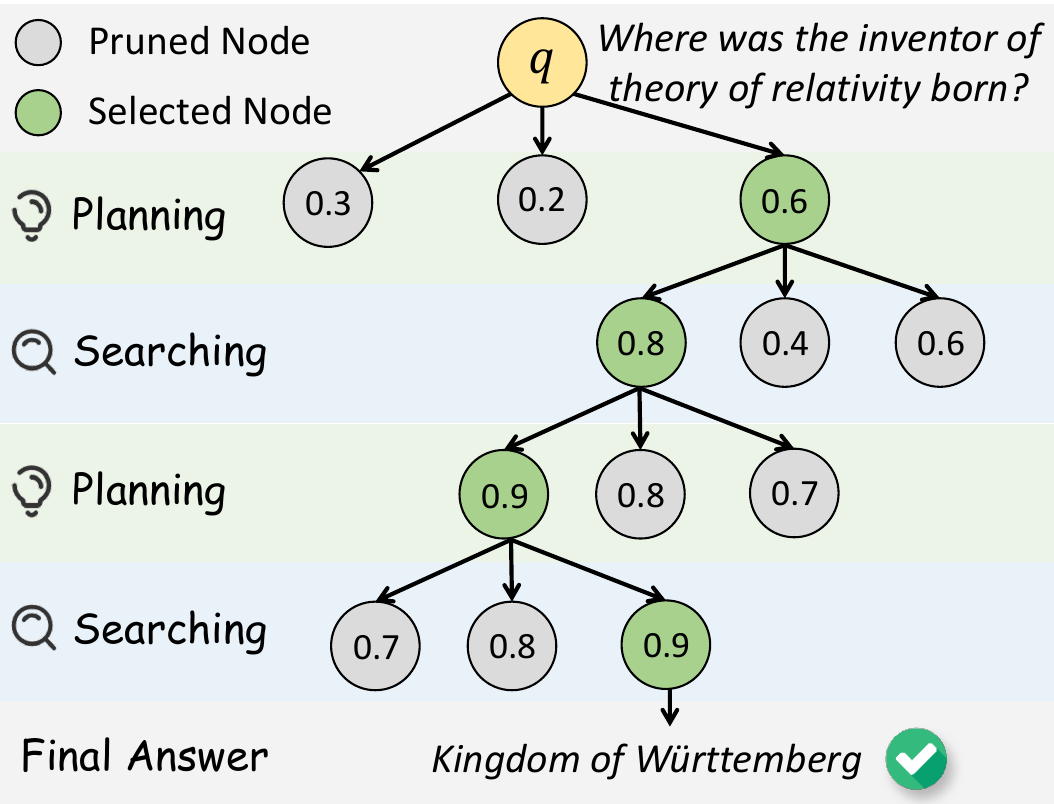}
    \caption{The illustration of Hierarchical Beam Search. During planning, the policy model generates and prunes candidate plans using the planning value model. During searching, it generates queries, retrieves documents, and prunes results using the search value model. This process iterates until the final answer is reached.}
\end{figure}

Although Agentic RAG shows superior performance, it faces several inherent challenges:
(1) The success of the reasoning process depends not only on high-quality planning but also on the accuracy of the retrieved information. While planning can be improved with high-quality training data and sophisticated pipelines, ensuring accurate retrieval remains challenging, as it depends on both the quality of the generated queries and the retrieval system’s performance.
(2) Evaluating the quality of each reasoning step is difficult due to the lack of explicit supervision signals. Most RAG datasets only provide final answers without feedback on intermediate steps, making it hard to assess and improve the quality of individual reasoning stages.
(3) The exponential candidate space for planning and searching creates a large, computationally intensive search space, making it challenging to efficiently identify optimal paths.

To address these challenges, we propose \textbf{\Ours}, a novel Agentic RAG framework that decouples planning and search processes using dual value models.
\textbf{To enhance the success probability of each reasoning step}, we introduce planning exploration and search exploration phases. The policy model generates multiple potential plans, which are evaluated by a planning value model to select the most promising options. Based on these plans, the policy model generates multiple queries to retrieve relevant documents. These search results are then ranked by the search value model to ensure the reliability of the retrieval process.
\textbf{To efficiently assess the quality of each reasoning step}, we introduce Monte Carlo Tree Search (MCTS) \cite{silver2017mastering} to guide the exploration of potential reasoning paths.
During MCTS simulations, the LLM acts as the judge to evaluate the quality of both the planning and search results, separately.
Through iterative MCTS simulations, the rewards derived from final answer correctness are back-propagated to update the LLM scores, refining the LLM's scores and correcting potential inaccuracies.
\textbf{To combat the exponential search space}, we propose pruning the planning and search spaces using a planning value model and a search value model. These models are trained on reward signals derived from the reasoning tree constructed through MCTS annotation. During inference, we employ Hierarchical Beam Search. At each step, the policy model generates multiple plans, which are evaluated by the planning value model to retain only the most promising ones. Based on these plans, the policy model generates search queries to retrieve relevant documents. The search value model then evaluates the retrieved results, preserving only the most valuable ones. This iterative process continues until either the maximum depth is reached or no further nodes can be expanded, ensuring effective reasoning.

Our contributions can be summarized as follows:
\begin{itemize}[topsep=1pt, partopsep=1pt, leftmargin=12pt, itemsep=-1pt]
    \item We introduce \Ours, a novel Agentic RAG framework that decouples planning-search processes with dual value models, enabling independent optimization of plan reasoning and search grounding.
    \item We propose improving the success rate of each step by fully exploring the planning and search spaces. We utilize MCTS to accurately assess planning and search quality, while Hierarchical Beam Search is employed to efficiently prune the exponential candidate space.
    \item Extensive experiments on five datasets across policy models of different parameter sizes demonstrate the effectiveness of our method.
\end{itemize}

%% file: Section/Background.tex
\section{Background}

\begin{figure*}[htbp]
    \centering
    \includegraphics[width=.99\linewidth]{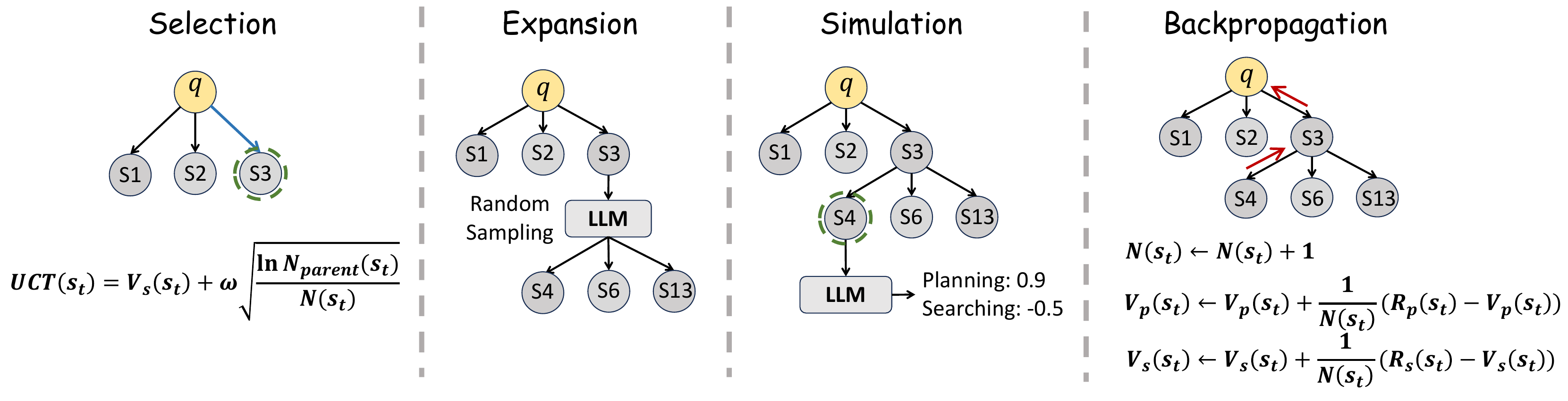}
    \caption{Single iteration of MCTS Annotation. The iteration is repeated until the maximum number of iterations is reached or no further nodes in the tree can be expanded.}
    \label{fig:mcts}
\end{figure*}

In Agentic RAG, given a user query \( q \), the policy model conducts multi-step reasoning and retrieves external knowledge to produce the final answer. Each step typically involves two stages: planning and search. 
In the planning stage, the policy model \(\mathcal{M}\) reasons based on the interaction history \(\tau_{t-1}\) and generates a plan \( p_t \):
\[
p_t = \mathcal{M}(\tau_{t-1}),
\]
where \(\tau_{t-1} = \{q, p_1, q_1, d_1, \dots, p_{t-1}, q_{t-1}, d_{t-1}\}\) represents the previous reasoning path.

In the search stage, the policy model generates search queries and retrieves external documents using an off-the-shelf search engine:
\begin{align}
    q_t &= \mathcal{M}(\tau_{t-1}, p_t), \\
    d_t &= \text{Retrieve}(q_t),
\end{align}
where \( d_t \) denotes the retrieved documents.

The success of each step depends on two key factors: the quality of the planning and the precision of the search. While planning can be improved through high-quality training data, search is subject to uncertainties due to challenges in query formulation and retriever performance.

To enhance the success rate of each step, we encourage the policy model to fully explore both the planning and search spaces through sampling. These sampled paths are then refined using a planning value model and a search value model, respectively, ensuring more accurate outcomes.

%% file: Section/Methodology.tex
\section{Approach}
\subsection{Overview}
\begin{figure*}[htbp]
    \centering
    \includegraphics[width=.99\linewidth]{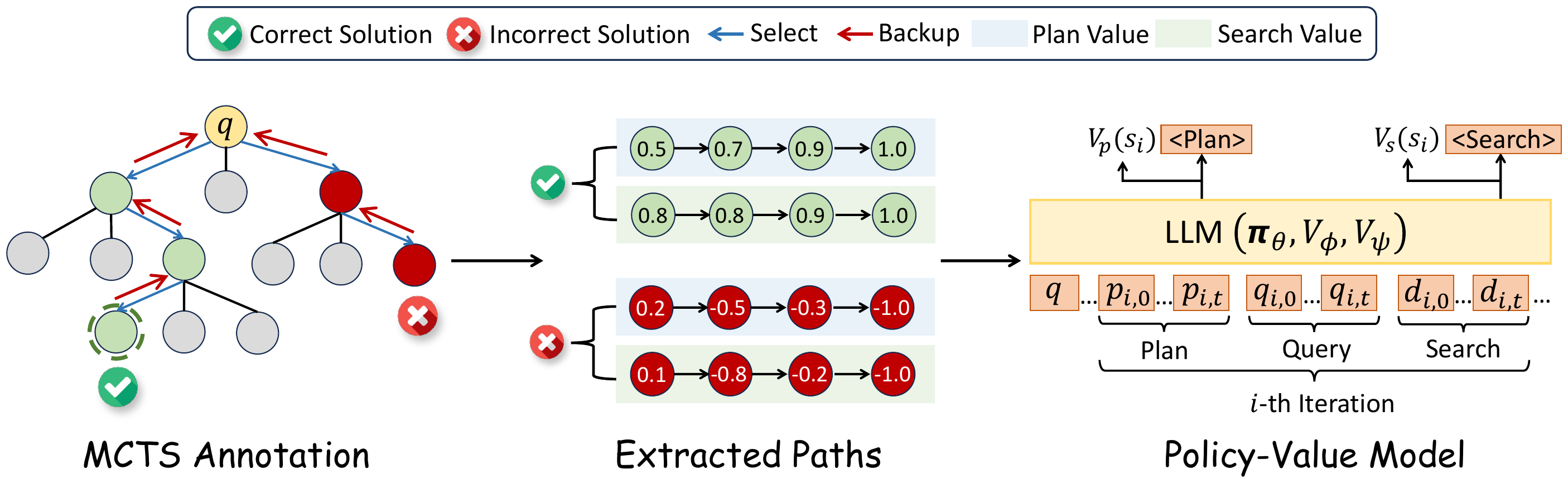}
    \caption{Overview of the proposed method: The MCTS algorithm constructs reasoning trees for training queries, from which correct and incorrect paths are extracted to train the policy and value models. During inference, Hierarchical Beam Search iteratively refines planning and search for accurate reasoning and retrieval.}
    \label{fig:framework}
\end{figure*}

\cref{fig:framework} presents an overview of our framework. The process begins with the application of Monte Carlo Tree Search (MCTS) to construct a reasoning tree for the queries in the training dataset. Each node in the tree represents a reasoning step, encompassing both planning and search results. From these trees, we extract both correct and incorrect paths, which are subsequently utilized to train the policy model and the value models, respectively. During the inference phase, we introduce a hierarchical beam search algorithm, where at each layer, the policy model fully explores the planning and search spaces, and the value models select the best candidate for further refinement.

\subsection{MCTS Annotation}
During MCTS annotation, we prompt the LLM to generate plans and search queries, interactively collaborating with the retriever to iteratively expand the reasoning tree. The process runs for multiple simulations and terminates when the maximum iteration number is reached, or no further paths can be expanded.
For the $i$-th simulation, MCTS conducts four operations to expand the tree:

\paragraph{Selection}
The \(i\)-th simulation begins with \(s_0\), representing the input query. The algorithm selects nodes according to the Upper Confidence Bound for Trees (UCT) criterion \cite{rosin2011multi}:

\begin{equation}
\small
{UCT}(s_t) = V_s(s_t) + w\sqrt{\frac{\ln{N_{parent}(s_t)}}{N(s_t)}}
\end{equation}
where \(V_s(s_t)\) represents the reward of the search result, and \(w\) controls the balance between exploration and exploitation.
The reason we choose \( V_s(s_t) \) to calculate the UCT score is that the quality of the search results serves as a reliable indicator of a step's potential to arrive at the correct answer.

\paragraph{Expansion}
After selecting the node to be expanded, the LLM generates the next plan and query based on the reasoning status. For simplicity, assume the chosen node \(s_t\) corresponds to the intermediate reasoning trajectory \(\tau_{t-1}\). The expansion process is as follows:
\begin{align}
    p_t, q_t &= \text{LLM}(\tau_{t-1}) \\
    d_t &= \text{Retrieve}(q_t)
\end{align}
To ensure diversity, we employ sampling generation with a higher temperature.

\paragraph{Simulation}
The simulation evaluates the quality of planning and search at each step and assigns reward values. For intermediate nodes, the LLM assesses the quality of planning and search, assigning a value between \(-1\) and \(1\), where \(1\) indicates high quality and \(-1\) indicates low quality:
\begin{align}
    R_p(s_t), R_s(s_t) = \text{LLM}(\tau_{t-1}, p_t, q_t)
\end{align}
For terminal nodes, if the final answer is correct, both planning and search rewards are set to \(1\); otherwise, they are set to \(-1\).

\paragraph{Backpropagation}
At the end of the \(i\)-th simulation, each edge along the path from the leaf node \(s_t\) to the root undergoes a backward pass update. The updates to their values and visiting counts are executed as follows:
\begin{equation}
\small
   \begin{aligned}
    N(s_t) &\leftarrow N(s_t)  + 1 \\
    V_{p}(s_t) &\leftarrow V_{p}(s_t) + \frac{1}{N(s_t)} (R_p(s_t)-V_{p}(s_t)) \\
    V_{s}(s_t) &\leftarrow V_{s}(s_t) + \frac{1}{N(s_t)} (R_s(s_t)-V_{s}(s_t))
\end{aligned} 
\end{equation}
This backpropagation process is crucial because it corrects potential inaccuracies in the LLM's generated scores using the answer correctness signal.

\subsection{Model Training}
In our framework, the policy model \(\pi_{\theta}\) is initialized with a pre-trained LLM. We extend this model to derive the planning value model \(V_\phi\) and search value model \(V_\psi\) by adding two auxiliary linear layers with a \(\texttt{Tanh}\) activation function. These layers operate alongside the traditional softmax layer responsible for token prediction, as illustrated in the rightmost panel of \cref{fig:framework}. This design ensures that the policy model and the value models share the majority of their parameters, promoting parameter efficiency and joint optimization.

To construct the training signals for the policy model and the value models, we sample solution paths from the tree constructed through multiple rounds of MCTS. These paths are denoted as \(\mathbf{x}^+\) (correct solutions) and \(\mathbf{x}^-\) (incorrect solutions). We then apply a multi-task loss function to jointly update all the models:
\begin{equation}
\small
\begin{split}
    \mathcal{L} =  - \log \pi_{\theta}(\mathbf{x}^+|\mathbf{q}) + \beta \cdot \sum_{t=1}^{T(\mathbf{x})} &\Big( \|V_\phi(\mathbf{s}_t) - {V}_p(\mathbf{s}_t) \|^2 \\ 
    + &\|V_\psi(\mathbf{s}_t) - {V}_s(\mathbf{s}_t) \|^2 \Big)
\end{split}
\end{equation}
Here, the first term represents the negative log-likelihood loss for next-token prediction in correct solutions, guiding the policy model to generate accurate predictions. The second term captures the loss in value prediction for both correct and incorrect solutions, ensuring the value models provide reliable estimates of expected rewards at each node. \(T(\mathbf{x})\) denotes the number of steps in the solution path \(\mathbf{x}\), and \(\beta\) is a tunable hyperparameter that controls the weight of the value loss term.

\subsection{Model Inference}
After obtaining the trained policy model, it can be directly used to conduct reasoning. However, this greedy decoding process fails to fully explore the planning and search spaces, limiting its ability to identify optimal reasoning paths. To address this issue, we propose a hierarchical beam search algorithm to encourage the policy model to thoroughly explore both the planning and search spaces.

\input{Tables/main_result}

\paragraph{Hierarchical Beam Search}
At each step, the policy model first samples multiple possible plans, which are ranked and filtered by the planning value model.
Based on the most promising plan, the policy model generates multiple search queries, which are used to retrieve relevant documents. The retrieved documents are then evaluated by the search value model to select the most valuable result.
This iterative process continues until the maximum depth is reached or no further paths can be expanded.
Finally, the answers are evaluated by the planning value model, and the answer with the highest value is selected as the output.
This approach ensures a more comprehensive exploration of both the planning and search spaces, leading to higher quality and more reliable results.

%% file: Tables/main_result.tex
\begin{table*}[t]
\centering
\resizebox{1.0\textwidth}{!}{
\begin{tabular}{lcccccccccccc}
\toprule
\multicolumn{1}{c}{\multirow{2.5}{*}{\textbf{Method}}} & \multicolumn{2}{c}{\textbf{HotpotQA}} & \multicolumn{2}{c}{\textbf{2WikiMulti}} & \multicolumn{2}{c}{\textbf{MuSiQue}} & \multicolumn{2}{c}{\textbf{Bamboogle}} & \multicolumn{2}{c}{\textbf{TriviaQA}} & \multicolumn{2}{c}{\textbf{AVG}} \\
\cmidrule(r){2-3} \cmidrule(r){4-5} \cmidrule(r){6-7} \cmidrule(r){8-9} \cmidrule(r){10-11} \cmidrule(r){12-13}
\multicolumn{1}{c}{} & \textbf{EM} & \textbf{F1} & \textbf{EM} & \textbf{F1} & \textbf{EM} & \textbf{F1} & \textbf{EM} & \textbf{F1} & \textbf{EM} & \textbf{F1} & \textbf{EM} & \textbf{F1} \\
\midrule
\headercolor
\multicolumn{13}{c}{\textbf{Qwen-2.5-7B-Instruct}} \\
Direct &  18.20 & 24.10 & 25.80 & 27.61 & 5.80 & 10.96 & 16.94 & 19.00 & 43.20 & 41.84 & 21.99 & 24.70  \\
CoT &  20.64 & 26.64 & 24.00 & 26.44 & 7.80 & 13.69 & 15.32 & 18.81 & 46.28 & 46.03 & 22.81 & 26.32    \\
Standard & 26.00 & 30.04 & 15.00 & 17.31 & 7.20 & 11.57 & 19.35 & 21.77 & 58.80 & 56.42 & 25.27 & 27.42   \\
\midrule
Iterative & 13.20 & 16.40 & 7.80 & 9.59 & 3.80 & 6.39 & 20.97 & 24.60 & 40.00 & 37.66 & 17.15 & 18.93   \\
Gen-Retrieve  & 24.40 & 28.15 & 18.80 & 19.68 & 8.20 & 12.95 & 19.35 & 21.41 & 55.20 & 52.93 & 25.19 & 27.02    \\
Judge-Retrieve &  25.80 & 30.88 & 17.40 & 19.96 & 7.60 & 12.18 & 19.35 & 21.77 & 55.20 & 52.92 & 25.07 & 27.54    \\
\midrule
RAgent &  26.40 & 30.95 & 26.00 & 28.70 & 9.00 & 14.57 & 30.65 & 35.01 & 57.20 & 50.89 & 29.85 & 32.02   \\
Search-o1 & 29.80 & 32.37 & 29.60 & 31.33 & 12.40 & 16.67 & 31.45 & 35.95 & 53.60 & 50.70 & 31.37 & 33.40   \\
GreedyAgent &  34.94 & 32.86 & 34.00 & 34.79 & 12.40 & 17.01 & 36.59 & 39.03 & 61.40 & 53.02 & 35.87 & 35.34    \\
\Ours &  \textbf{38.62} & \textbf{36.60} & \textbf{35.87} & \textbf{35.03} & \textbf{17.20} & \textbf{17.73} & \textbf{42.28} & \textbf{46.52} & \textbf{65.66} & \textbf{58.08} & \textbf{39.93} & \textbf{38.79}   \\
\midrule
\headercolor
\multicolumn{13}{c}{\textbf{Qwen-2.5-14B-Instruct}} \\
Direct &  22.00 & 27.37 & 25.60 & 27.24 & 6.20 & 12.63 & 12.90 & 15.85 & 54.00 & 51.47 & 24.14 & 26.91   \\
CoT &  26.00 & 30.98 & 25.60 & 27.52 & 9.40 & 15.30 & 33.06 & 37.58 & 60.40 & 58.82 & 30.89 & 34.04    \\
Standard &  27.40 & 28.51 & 34.00 & 20.75 & 9.40 & 12.91 & 16.94 & 19.95 & 60.40 & 54.70 & 29.63 & 27.36    \\
\midrule
Iterative &  15.00 & 16.24 & 5.40 & 6.74 & 5.21 & 8.70 & 10.48 & 15.16 & 42.20 & 36.30 & 15.66 & 16.63   \\
Gen-Retrieve & 26.80 & 27.10 & 33.20 & 21.23 & 8.40 & 11.70 & 19.35 & 21.41 & 61.60 & 57.30 & 29.87 & 27.75    \\
Jud-Retrieve &  27.40 & 28.38 & 33.40 & 20.16 & 9.00 & 11.49 & 18.55 & 21.11 & 60.40 & 55.32 & 29.75 & 27.29   \\
\midrule
RAgent &  37.40 & 38.18 & 33.87 & 34.19 & 16.20 & 18.72 & 37.90 & 43.28 & 65.80 & 59.92 & 38.23 & 38.86    \\
Search-o1 & 36.80 & 37.18 & 34.20 & 35.86 & 16.40 & 20.54 & 35.48 & 43.41 & 64.40 & 60.46 & 37.46 & 39.49    \\
GreedyAgent & 38.96 & 37.31 & 37.80 & 36.15 & 16.06 & 19.39 & 45.53 & 48.18 & 63.45 & 57.13 & 40.36 & 39.63    \\
\Ours & \textbf{43.35} & \textbf{39.56} & \textbf{41.84} & \textbf{38.37} & \textbf{18.38} & \textbf{21.82} & \textbf{47.15} & \textbf{49.77} & \textbf{72.44} & \textbf{62.98} & \textbf{44.63} & \textbf{42.50}    \\
\bottomrule
\end{tabular}
}
\caption{Evaluation results on five representative QA tasks. The \textbf{bold} fonts denote the best results in each dataset.}
\label{tab:main-result}
\end{table*}

%% file: Section/Experiment.tex
\section{Experiments}

\subsection{Datasets and Metrics}
We conduct experiments on five datasets spanning both single-hop and multi-hop question-answering (QA) tasks. Specifically, the multi-hop QA tasks include the 2WikiMultiHopQA dataset~\cite{2wikimultihopqa-ho-2020}, the HotpotQA dataset~\cite{yang2018hotpotqa}, the Bamboogle dataset~\cite{press2022measuring}, and the MuSiQue dataset~\citep{trivedi2022musique}, while the single-hop QA task is represented by the TriviaQA dataset~\cite{joshi2017triviaqa}.

To evaluate performance, we employ two key metrics: Exact Match (EM) and F1 Score. Under the EM metric, a predicted answer is deemed correct if its normalized form exactly matches any of the normalized versions of the reference answers in the provided answer list. The F1 score, on the other hand, quantifies the word-level overlap between the normalized predicted answer and the reference answers, providing a measure of the answer's precision and recall.

\subsection{Baselines}
We compare \Ours with the following three categories of methods:
\paragraph{Vanilla Prompting Methods}
This category includes direct prompting, Chain-of-Thought (CoT), and standard Retrieval-Augmented Generation (RAG). Direct prompting instructs the model to generate answers directly without retrieving external resources. Chain-of-Thought guides the model to reason step by step before arriving at the final answer. Standard RAG first retrieves relevant documents from an external corpus and then generates the answer based on the retrieved information.

\paragraph{Advanced RAG Methods}
This category includes Iterative RAG~\citep{searchain}, Judge-then-retrieve~\citep{asai2023self}, and Generate-then-retrieve~\citep{query2doc}. We implement all these baselines in our experiments. Iterative RAG decomposes the query into sub-queries, retrieves and generates answers for each, and then combines them to produce the final answer. Judge-then-retrieve first determines whether retrieval is necessary and then generates the final answer using either internal knowledge or retrieved documents. Generate-then-retrieve directly generates an answer, concatenates the answer with the question, and then retrieves and generates a refined answer.

\paragraph{Agentic RAG Methods}
This category includes RAgent, Search-o1~\cite{li2025search}, and GreedyAgent. These methods operate by iteratively searching for the necessary information to answer the question. At each step, the policy model autonomously decides when and what to retrieve. Search-o1 enhances this approach by incorporating a Reason-in-Document module, which condenses retrieved documents into reasoning steps while preserving the logical flow of the reasoning chain.
GreedyAgent is a greedy variant of \Ours, whose beam size is set to 1.

\subsection{Implementation Details}
To demonstrate the generality of our method, we initialize the policy with two large language models (LLMs) of different parameter sizes: Qwen2.5-7B-Instruct\footnote{\tiny \url{https://huggingface.co/Qwen/Qwen2.5-7B-Instruct}}\cite{qwen2.5} and Qwen2.5-14B-Instruct\footnote{\tiny \url{https://huggingface.co/Qwen/Qwen2.5-14B-Instruct}}\cite{qwen2.5}. During Monte Carlo Tree Search (MCTS) annotation, we employ Qwen-Turbo to predict the next action and evaluate the scores for planning and searching.  
The policy model and value models are fine-tuned over 10 epochs with a batch size of 4 and a learning rate of 1e-6, utilizing 8 NVIDIA A100 80GB GPUs. For retrieval, we use the Wikipedia dump from January 27, 2020, as our corpus and employ DPR~\cite{karpukhin-etal-2020-dense} as our dense retriever. For each query, we retrieve the top-5 most relevant documents from the retrieval corpus.  
Additional implementation details can be found in \cref{sec:train_details}.
To promote reproducibility, we plan to open-source the code upon acceptance of this work.

\subsection{Main Results}
In this section, we present the results of experiments conducted on five QA datasets using two model backbones, respectively. Based on the results in \cref{tab:main-result}, several observations can be made:

First, our method achieves superior performance on all datasets across different policy models, verifying the effectiveness of our approach. Notably, when using Qwen2.5-7B-Instruct as the policy model, \Ours achieves a 25.8\% relative average improvement over the best-performing baseline. This improvement is attributed to the application of hierarchical beam search, which enables the policy model to thoroughly explore the planning and search spaces, significantly increasing the likelihood of identifying the correct path.

Second, among the baselines, agentic RAG methods outperform both prompting methods and advanced RAG methods. This is primarily due to the flexibility agentic RAG provides, allowing the policy model to dynamically decide what to retrieve and when to retrieve. This capability is especially important for complex queries requiring multi-step reasoning, as demonstrated by strong performance on multi-hop datasets such as Bamboogle.

Third, when comparing policy models of different sizes, larger models (e.g., Qwen2.5-14B-Instruct) generally yield better performance, as expected, due to their higher model capacity. However, after applying Hierarchical Beam Search (HBS), the performance of \Ours with the 7B policy model becomes comparable to that of the 14B model, highlighting the potential for smaller models to achieve competitive performance through inference-time scaling techniques.

\section{Analysis}
\subsection{Ablation Study}
\input{Tables/ablation}
In this section, we analyze the effectiveness of planning expansion and search expansion by removing these components and observing the resulting performance changes, as shown in \cref{tab:ablation}. 

The results demonstrate that removing either planning expansion or search expansion leads to a decline in performance, underscoring the importance of thoroughly exploring both the planning space and the search space. Notably, the removal of planning expansion results in a more significant performance drop. This is because the planning typically defines the search space; if the plan is suboptimal, it becomes challenging to retrieve high-quality results. Therefore, planning expansion plays a more critical role in ensuring robust model performance.

\subsection{Scaling with Planning and Searching}
\input{Tables/hyper}
During inference, we employ hierarchical beam search, which involves two key hyperparameters: the planning expansion size \( B_1 \) and the search expansion size \( B_2 \). To investigate their impact on model performance, we conduct experiments on the HotpotQA, 2WikiMultiHopQA, and MuSiQue datasets, varying these parameters within the range of 1 to 5. Based on the results presented in \cref{tab:hyper}, several observations can be made.

First, for the planning expansion size, model performance peaks when the expansion size is set to approximately 3. Values smaller or larger than this threshold result in a decline in performance. This is primarily because a larger planning expansion size provides the model with more opportunities to identify the optimal plan. However, when the expansion size becomes too large, the planning value model struggles to effectively rank and select the best plan due to increased complexity.

Second, \textbf{for the search expansion size, we observe that larger expansion sizes generally lead to improved performance}. This is because a larger search expansion size increases the likelihood of retrieving optimal evidence that can lead to the correct answer. Compared to the planning value model, the search value model faces relatively less difficulty in ranking search results, as it can directly evaluate the retrieved evidence, whereas the planning value model must rely on complex patterns learned from training data to make decisions.

\subsection{Effectiveness of Value Models}
\begin{figure}[tb]
\includegraphics[width=0.99\linewidth]{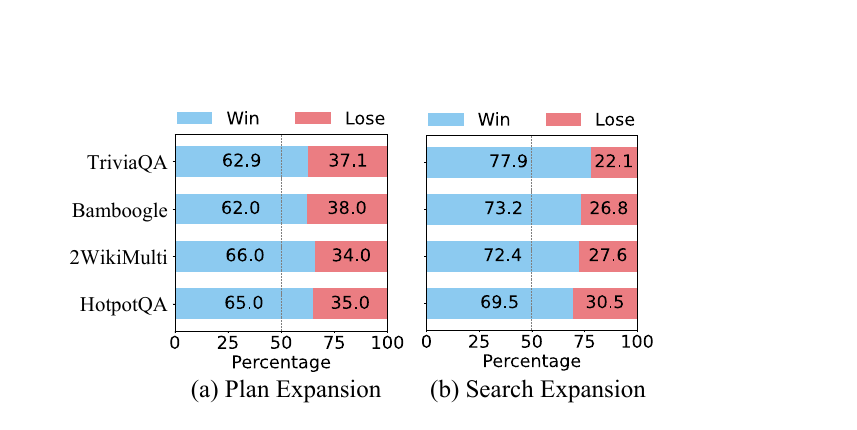}
\centering
\caption{We analyze the effectiveness of value models by replacing the value ranking with random sampling.}
\label{fig:win_lose}
\vspace{-1em}
\end{figure}
In this section, we analyze the accuracy of the planning value model and the search value model. Specifically, during the beam ranking stage, instead of using our value model to rank, we randomly select one plan or search result and compare the performance of random selection with ranking by the value model. Based on the results shown in \cref{fig:win_lose}, several observations can be made:

First, for both planning expansion and search expansion, ranking by the learned value model achieves better performance compared to random selection. This verifies that both value models can accurately measure the quality of plans and search results. 
Second, the performance superiority is more pronounced for search expansion. \textbf{This is because determining the value of search results is relatively easier than evaluating plans}. Typically, the value of a search result can be directly assessed by checking whether it contains the answer to the search query. In contrast, evaluating the quality of a plan is more challenging, as there are no obvious patterns to determine its effectiveness. 
Therefore, when computational resources are constrained, allocating more resources to search expansion may be a more robust strategy.

\subsection{Case Study}
\begin{figure*}[htbp]
    \centering
    \includegraphics[width=.98\linewidth]{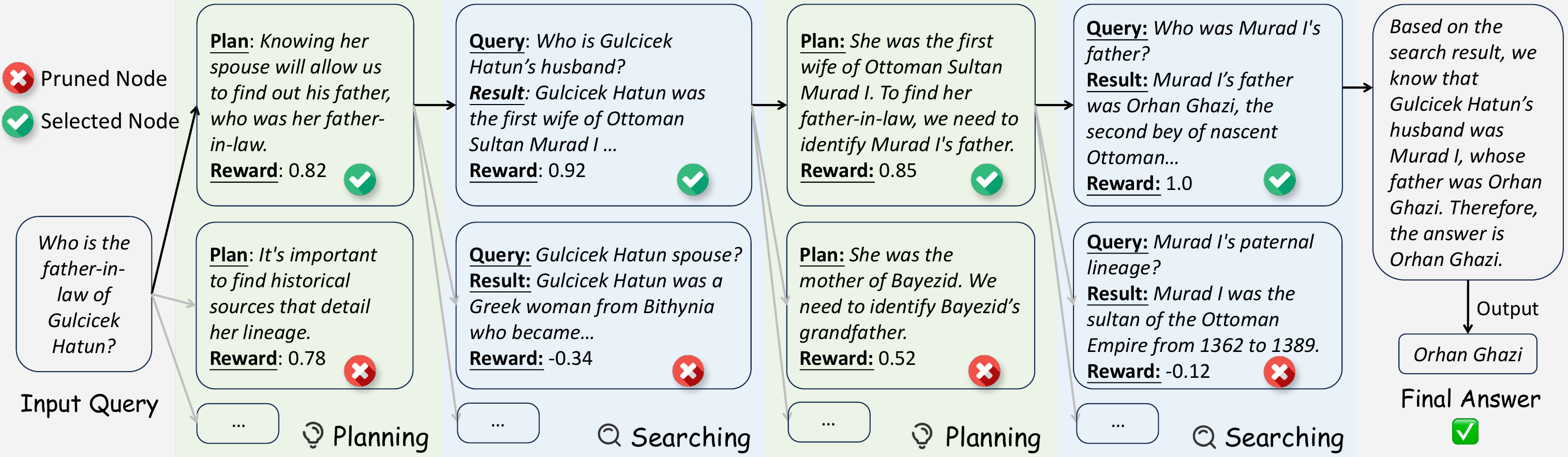}
    \caption{Case Study of \Ours. The example illustrates how \Ours optimizes reasoning by decoupling planning and search using hierarchical reward modeling. The reasoning tree dynamically selects high-reward paths while pruning suboptimal ones, demonstrating the effectiveness of Hierarchical Beam Search.}
    \label{fig:case}
\end{figure*}
In this section, we present a case study from the MuSiQue dataset in \cref{fig:case}.

Given the query ``Who is the father-in-law of Gulcicek Hatun?'', the policy model generates plans, such as searching for Gulcicek Hatun's spouse or lineage. The planning value model assigns a higher reward to the spouse search plan, pruning the lower-value alternative.
The policy model then creates search queries like ``Who is Gulcicek Hatun's husband?'' and ``Gulcicek Hatun Spouse''. The first query retrieves direct information about her husband, Murad I, and receives a high positive reward, while the less relevant result is pruned.
Next, the policy model searches for Murad I's father, generating queries like ``Who was Murad I's father?'' The result identifying Orhan Ghazi as his father receives a high reward, while irrelevant results are pruned. The final answer, Orhan Ghazi, is output.
This case study illustrates how planning and search expansion broaden the candidate space, while the value models identify the most valuable candidates, validating our framework's effectiveness in handling complex queries.

%% file: Tables/ablation.tex
\begin{table}[t]
\centering
\resizebox{\linewidth}{!}{
\begin{tabular}{lcccc}
\toprule
 \multirow{2}{*}{Methods} & \multicolumn{2}{c}{HotpotQA} & 
\multicolumn{2}{c}{TriviaQA}   \\
\cmidrule(lr){2-5} 
& EM & F1 & EM & F1  \\
\midrule
\Ours &  38.62 & 36.60 & 65.66 & 58.08  \\
-w/o Planning & 35.35 & 32.84 & 62.10 & 54.78     \\
-w/o Searching &  36.75 & 35.05 & 62.58 & 55.63   \\
-w/o Both & 34.94 & 32.86 & 61.40 & 53.02   \\
\bottomrule
\end{tabular}
}
\caption{Ablation Study. We experiment by gradually removing all model components.}
\label{tab:ablation}
\end{table}

%% file: Tables/hyper.tex
\begin{table}[t]
\centering
\resizebox{\linewidth}{!}{
\begin{tabular}{ccccccc}
\toprule
 \multirow{2}{*}{Size} & \multicolumn{2}{c}{HotpotQA} & 
\multicolumn{2}{c}{2WikiMulti}   & 
\multicolumn{2}{c}{MuSiQue}   \\
\cmidrule(lr){2-3} \cmidrule(lr){4-5} \cmidrule(lr){6-7} 
& EM & F1 & EM & F1  & EM & F1  \\
\midrule
\headercolor
\multicolumn{7}{c}{Plan Expansion Size} \\
1 &  35.35 & 32.84 & 34.74 & 34.68 & 14.20 & 16.66  \\
2 &  37.83 & 37.48 & 36.40 & 35.68 & 15.43 & 18.53      \\
3 &  38.62 & 36.60 & 35.87 & 35.03 & 17.20 & 17.73     \\
4 & 37.78 & 36.00 & 35.60 & 36.14 & 15.23 & 19.20      \\
5 & 35.29 & 34.62 & 33.87 & 34.19 & 14.63 & 18.42       \\
\midrule
\headercolor
\multicolumn{7}{c}{Search Expansion Size} \\
1 &  36.75 & 35.05 & 33.40 & 34.04 & 14.80 & 17.78    \\
2 & 38.87 & 37.68 & 34.07 & 33.85 & 15.03 & 18.19     \\
3 & 38.18 & 36.85 & 33.87 & 34.38 & 16.23 & 18.92     \\
4 &  37.78 & 35.36 & 35.34 & 34.79 & 15.43 & 17.78   \\
5 & 40.20 & 37.46 & 35.34 & 35.99 & 17.60 & 19.59    \\
\bottomrule
\end{tabular}
}
\caption{We vary the expansion sizes within the range of 1 to 5 and observe the performance changes.}
\vspace{-1em}
\label{tab:hyper}
\end{table}

%% file: Section/Related_work.tex
\section{Related Work}
\subsection{Agentic Retrieval-Augmented Generation}
Despite significant advancements, Large Language Models (LLMs) often generate responses that include hallucinated facts and inaccurate information \cite{ji2023survey, shuster2021retrieval, zhang2023sac3}, which compromises their reliability and limits their practical applicability in real-world scenarios.
To mitigate this issue, researchers have turned to Retrieval-Augmented Generation (RAG), which integrates external knowledge to improve the accuracy of responses \cite{ram2023context, shi2023replug, rashkin2021measuring, gao2022rarr, bohnet2022attributed, menick2022teaching,chen2024learning}. 
By dynamically retrieving information from external documents, RAG enables LLMs to ground their outputs in verifiable evidence.

While RAG offers substantial improvements, it remains limited by its reliance on static workflows.
Agentic RAG presents a more promising approach by incorporating agents into the RAG pipeline. For instance, Self-RAG \cite{asai2023self} employs a self-reflection mechanism to iteratively predict reflection tokens during training. Auto-RAG \cite{yu2024auto} systematically plans retrievals and refines queries to acquire valuable knowledge through multi-turn iterations. MindSearch \cite{chen2024mindsearch} mimics human cognitive processes in web information seeking and integrates them with an LLM-based multi-agent framework. PlanxRAG \cite{verma2024plan} isolates the reasoning plan as a directed acyclic graph (DAG) outside the LM's working memory. Search-o1 \cite{li2025search} incorporates an agentic search process into reasoning, allowing for the dynamic retrieval of information whenever LLMs face uncertain knowledge points.

\subsection{Enhancing LLMs with Search}
The application of search techniques to enhance LLMs has garnered considerable attention \cite{yao2023tree}. Numerous studies have demonstrated that MCTS can significantly improve the reasoning capabilities of LLMs by generating diverse reasoning paths. For example, AlphaMATH \cite{chen2024alphamath} utilizes MCTS to eliminate the need for process annotations from humans or GPTs. Similarly, SVPO \cite{chen2024step} employs MCTS to automatically annotate step-level preferences for multi-step reasoning. Llama-berry \cite{zhang2024llama} leverages MCTS to facilitate more efficient exploration of solution spaces.

Other notable works include CoAT \cite{pan2025coat}, which integrates MCTS with associative memory for structured reasoning, and MCTSr \cite{zhang2024accessing}, which applies MCTS to self-refine mathematical solutions through tree-search iterations. AirRAG \cite{feng2025airrag} activates intrinsic reasoning capabilities and expands the solution space for specific tasks using MCTS.

%% file: Section/Conclusion.tex
\section{Conclusion}
In this paper, we propose \Ours, a novel framework that decouples planning and search processes using dual value models, enabling independent optimization of plan reasoning and search grounding. Our approach constructs a reasoning tree, where each node represents planning and search steps. We leverage MCTS to efficiently assess the quality of each step. During inference, hierarchical beam search iteratively refines planning and search candidates through reward-guided optimization. Extensive experiments across policy models of varying parameter sizes demonstrate the effectiveness of our method.

%% file: Section/Limitations.tex
\section*{Limitations}
In this paper, we propose an agentic RAG framework that fully explores the planning and search spaces. We acknowledge two limitations of our method. First, the MCTS annotation process requires multiple simulations, which can lead to additional labeling costs.
Second, our current approach focuses on retaining only the single most promising plan and search result at each step. The exploration of retaining multiple promising plans and search results is left for future work.

%% file: Section/Ethics.tex
\section*{Ethics Statement}
This work complies with the ACL Ethics Policy. All datasets and LLMs used are publicly available. Our research focuses on improving the performance of agentic RAG, and we do not anticipate any negative ethical impacts.

%% file: Section/Appendix.tex
\section{Dataset Statistics}
The dataset statistics used in this paper are shown in \cref{tab:datasets}.
\input{Tables/dataset_statistics}

\section{Implementation Details}
\label{sec:train_details}
\paragraph{Dataset Construction}
We sample queries from the training data of NQ, HotpotQA, 2WikiMultiHopQA and MuSiQue datasets and conduct the MCTS annotations.
During MCTS annotation,
following \citet{chen2024alphamath},  the parameters \( w, \beta \) are set to 1.4 and 0.1, respectively.
The maximum number of iterations is configured to 20.
We employ Qwen-Turbo to predict the next action and evaluate the scores for planning and searching.
We then sample correct paths and incorrect paths from the constructed trees.
The correct paths are used to train the policy model, while both paths are used to train the value models.

\paragraph{Training Process}
The policy model and value models are fine-tuned over 10 epochs with a batch size of 4 and a learning rate of 1e-6, utilizing 8 NVIDIA A100 80GB GPUs.

\paragraph{Inference Process}
During inference, both the plan expansion size and the search expansion size are set to 3.

\paragraph{Training Cost}
\input{Tables/training_cost}
We conduct training using 8 NVIDIA A100 80GB GPUs, with the detailed time costs for both labeling and training summarized in \Cref{tab:training-time}.
As shown, the overall training cost is reasonable, particularly since convergence is typically achieved in about half of the total training time.
Moreover, MCTS is used only once during offline labeling and introduces no additional overhead during inference.

\section{Compare with Non-Agentic Baselines}
\input{Tables/compare_non_agentic_baselines}
To ensure a fair comparison, we conducted additional experiments comparing our method with Self-RAG~\cite{asai2023self} and Adaptive-RAG~\cite{jeong2024adaptive} using their official open-source implementations. Specifically, we employed the official open-sourced Self-RAG model, and used Qwen2.5-7B-Instruct as the backbone model for both \Ours and Adaptive-RAG.
The results are presented in \Cref{tab:non-agentic}. \Ours outperforms both Self-RAG and Adaptive-RAG across all datasets, demonstrating the effectiveness of our agentic reasoning framework.

\section{Prompts}
The prompts used in MCTS annotations are listed below:
\input{Tables/prompts}

%% file: Tables/dataset_statistics.tex
\begin{table*}[h]
\centering
\resizebox{\linewidth}{!}{
\begin{tabular}{lccccc}
\toprule
\textbf{Settings} & \textbf{TriviaQA} & \textbf{Bamboogle} & \textbf{HotpotQA} & \textbf{2WikiMultiHopQA} & \textbf{MuSiQue} \\
& \cite{joshi2017triviaqa}  & \cite{press2022measuring} & \cite{yang2018hotpotqa} & \cite{2wikimultihopqa-ho-2020} & \cite{trivedi2022musique} \\
\midrule
\multicolumn{6}{c}{\emph{Dataset statistics}} \\
Task & Single-Hop QA & Multi-Hop QA & Multi-Hop QA & Multi-Hop QA & Multi-Hop QA  \\
Test Data & 500 & 125 & 500 & 500 & 500 \\
\midrule
\multicolumn{6}{c}{\emph{Evaluation settings}} \\
Metrics & EM, F1  & EM, F1  & EM, F1  & EM, F1 & EM, F1 \\
\midrule
\multicolumn{6}{c}{\emph{Retrieval settings}} \\
Corpus & Wikipedia & Wikipedia & Wikipedia & Wikipedia & Wikipedia \\
Retriever & DPR & DPR  & DPR & DPR & DPR  \\
\bottomrule
\end{tabular}
}
\caption{Statistics and experimental settings of different tasks/datasets.}
\label{tab:datasets}
\end{table*}

%% file: Tables/training_cost.tex
\begin{table}[h]
\centering
\begin{tabular}{lccc}
\toprule
\textbf{Model} & \textbf{Stage} & \textbf{Total Time} & \textbf{Time to Converge} \\
\midrule
MCTS Labeling & Offline & 18h 31m & None \\
Qwen-2.5-7B-Instruct & Training & 12h 30m & \textasciitilde6h \\
Qwen-2.5-14B-Instruct & Training & 1d 2h & \textasciitilde12h \\
\bottomrule
\end{tabular}
\caption{Training and labeling time for different models.}
\label{tab:training-time}
\end{table}

%% file: Tables/compare_non_agentic_baselines.tex
\begin{table*}[t]
\centering
\resizebox{1.0\textwidth}{!}{
\begin{tabular}{lcccccccccccc}
\toprule
\multicolumn{1}{c}{\multirow{2.5}{*}{\textbf{Method}}} & \multicolumn{2}{c}{\textbf{HotpotQA}} & \multicolumn{2}{c}{\textbf{2WikiMulti}} & \multicolumn{2}{c}{\textbf{MuSiQue}} & \multicolumn{2}{c}{\textbf{Bamboogle}} & \multicolumn{2}{c}{\textbf{TriviaQA}} & \multicolumn{2}{c}{\textbf{AVG}} \\
\cmidrule(r){2-3} \cmidrule(r){4-5} \cmidrule(r){6-7} \cmidrule(r){8-9} \cmidrule(r){10-11} \cmidrule(r){12-13}
\multicolumn{1}{c}{} & \textbf{EM} & \textbf{F1} & \textbf{EM} & \textbf{F1} & \textbf{EM} & \textbf{F1} & \textbf{EM} & \textbf{F1} & \textbf{EM} & \textbf{F1} & \textbf{EM} & \textbf{F1} \\
\midrule
Self-RAG & 23.40 & 26.97 & 23.00 & 24.63 & 6.60 & 7.00 & 8.87 & 10.31 & 53.60 & 52.67 & 23.09 & 24.32 \\
Adaptive-RAG & 34.40 & 31.80 & 34.00 & 34.20 & 11.80 & 12.47 & 27.42 & 30.39 & 58.00 & 55.82 & 33.12 & 32.94 \\
\Ours & \textbf{38.62} & \textbf{36.60} & \textbf{35.87} & \textbf{35.03} & \textbf{17.20} & \textbf{17.73} & \textbf{42.28} & \textbf{46.52} & \textbf{65.66} & \textbf{58.08} & \textbf{39.93} & \textbf{38.79} \\
\bottomrule
\end{tabular}
}
\caption{Evaluation results on five representative QA tasks. The \textbf{bold} fonts denote the best results in each dataset.}
\label{tab:non-agentic}
\end{table*}

%% file: Tables/prompts.tex
\begin{figure*}[htbp]
\begin{prompt}[title={LLM Sample Prompt}, label=prompt:search]
**You are a highly capable web agent. Your task is to engage in multi-step reasoning and propose plans to reach a final answer for the given question.**\\

For each step, please include the following elements:\\

**Thought:** Offer a comprehensive and detailed analysis. This section should cover:\\
    - An analysis of the specific information required to address the question effectively and the information currently available.\\
    - If the information is enough to answer the question, you should conduct deep analysis based on the information and then answer the question.\\
    - If the information is not enough to answer the question, you should analyze whether the current plan progresses well. \\
        - If yes, predict the next action.\\
        - If no, reflect on why the progress is not good and then propose a new plan.\\

**Action:** Provide the next action. This section should cover:\\
   - If the information is enough to answer the question, you should output the final answer in format of Finish(put the answer here) without extra content. \\
   - If the information is not enough to answer the question, you should clearly specify the exact query for the next search in the format Search([List of Queries]) without extra content. Ensure the queries convey the same semantic information but are expressed differently to enhance the likelihood of finding the necessary information.\\

For the question: {query}, here is the reasoning process so far:\\
{history}\\

**The Output Format:**\\
- **Thought:** [Detailed analysis of the needed information, existing information, identifies whether information is enough. If enough, conduct analysis to obtain the final answer, else, identify what still needs to be searched]\\
- **Action:** [Finish(put the answer here) or Search([List of Queries])]\\

Please provide the plan for the next step:\\
\end{prompt}
\end{figure*}

\begin{figure*}[htbp]
\begin{prompt}[title={LLM Evaluation Prompt}, label=prompt:search]
**Task:**  Assess the effectiveness of the thought and the search result in the last reasoning step.\\
As an advanced web search agent, your role is to systematically evaluate the current step.\\
For the question: {query}, here is the reasoning process so far:\\
{history}\\

As an expert in web search, your tasks are as follows:\\
1. Analyze the thought in the last step: Evaluate the thought and determine its effectiveness in reaching the final answer. Assign a score between -1 and 1, where -1 means the thought is useless and 1 means the thought is very effective.\\
2. Analyze the search result in the last step: Evaluate the search result and determine its effectiveness in reaching the final answer. Assign a score between -1 and 1, where -1 means the search result was ineffective, and 1 means the search results were highly useful.\\

You should output the following elements\\
**Analysis of the thought:**\\
- Analyze whether the thought from the last step were helpful in progressing toward the final answer.\\
- Assign a score between -1 and 1, where -1 means the step was ineffective, and 1 indicates high usefulness.\\
- You must conclude the analysis with the format of "the value of the thought is ***x***", where x represent the value and * is the identifier. Remember that you must output the value x with identifier ***.\\

**Analysis of the search result:**\\
- Analyze whether the search query and search results from the last step were helpful in progressing toward the final answer.\\
- Assign a score between -1 and 1, where -1 means the step was ineffective, and 1 indicates high usefulness.\\
- You must conclude the analysis with the format of "the value of the search result is ***x***", where x represent the value and * is the identifier. Remember that you must output the value x with identifier ***.\\

Please begin by analyzing the previous step:
**Analysis of the thought:**\\
\end{prompt}
\end{figure*}